\documentstyle[twocolumn,prl,aps,epsf]{revtex}

\begin{document}                                                             

\title{Higgs production at Hadron colliders as a
probe of new physics}  
\author{ J.~Lorenzo~Diaz-Cruz \\
Theory Group, Lawrence Berkeley Laboratory\\
Berkeley, CA, 92710 \\
Instituto de Fisica, BUAP, 72570 Puebla, Pue, Mexico
}
\date{\today}
\maketitle
\begin{abstract}
\hspace*{-0.35cm}
The coupling of the Higgs with gauge bosons 
$g_{hVV}$  ($V=g,\gamma,W,Z$)  can receive non-decoupling corrections 
due to heavy quanta. Deviations from the SM predictions
are described by a set of parameters $\epsilon_V$, which can 
be related to the parameters $S,T$ for specific models.   
The Higgs production by gluon fusion 
can be used to probe $\epsilon_g$ at Tevatron, whereas the promising
Higgs decay into $ WW^*$ can probe $\epsilon_{W}$. 
We find that the resulting bounds may imply the exclusion of heavy 
particles that receive their mass directly from the SM Higgs, including
additional  standard or mirror families, chiral colored sextets and octet
quarks. 
Within the MSSM, we also find that gluon fusion is a sensitive
probe for the spectrum of squarks masses. 
\\[0.1cm]
\end{abstract}



\vspace*{-0.8cm}

\bigskip

{\bf {1.- Introduction.}}
The discovery of the Higgs boson as the remmant of the mechanism of
electro-weak symmetry breaking (EWSB), is one of the most cherised
goals of present and future high-energy experiments.
Within the minimal standard model (SM), the mass of the
physical Higgs particle is a free-parameter, but present data seems 
to favor a moderate mass ($105 < m_h < 220$ GeV) \cite{Higgsrev},  
however this conclusion may be changed by the presence
of  new phyics with an scale $\Lambda \simeq O(1) $ TeV
\cite{Meandthem}.  On the other hand, the minimal supersymmetric 
version of the SM (MSSM), which is one of its most appealing extensions,
predicts a light Higgs boson, with
an upper mass bound of about  130 GeV \cite{carenaetal}.
Detection of the full spectrum of Higgs bosons in the SM and beyond,
has been extensively studied, because it constitutes 
an important test of the possible realization of the Higgs mechanism
in fundamental physics.

 The characteristic Higgs boson couplings determine the
strategies employed for its search
at present and future colliders. For instance, at LEP 
the large couplings of the Higgs with massive gauge 
bosons allows to use the reaction $e^+e^- \to Z+h$;
whereas at next linear colliders (NLC) it will also be possible to
study Higgs production by WW and ZZ fusion. 
On the other hand, Hadron colliders can also test
these couplings, either through the reactions 
$pp \to W+h, Z+h$, or through the decays $h\to WW,ZZ$;
vector fusion can also be used for heavy Higgs masses.
The couplings of the Higgs with the heavier fermions, can be studied
either by open production of $t\bar{t}h, b\bar{b}h$ or by the loop-induced coupling
with gluon and photon pairs ($hgg, h\gamma\gamma$) \cite{hixhunter}. 
Any additional heavy particle that receive its
mass from the SM Higgs mechanism, will
couple to the Higgs with strength proportional to the particle 
mass itself, which means that such heavy quanta
will induce non-decoupling contributions to the
1-loop vertices $hgg,\,  h\gamma\gamma$.
Moreover, since these effects will compete at the same order
with the SM loop, it can be a significan effect that may
be probed at future colliders,  as has been
explored in previous studies \cite{chanofourth}. However
there is another important effect not considered previously,
namely that the presence of these new particles could
also induce non-decoupling corrections to the tree vertices
$hff$, $hWW$ and $hZZ$ \cite{veltmanetal}, which can affect the decay rate
of detectable signatures, and thus  must be included in the analysis.

In this papaer we study the effect of
additional colored particles in the loop vertex $hgg$,
and the bounds that could be obtained at Tevatron (RUN-II). 
We shall identify two cases, in the first (Scenario-I) we consider
the situation when the mass of heavy colored particles comes 
from the SM Higgs boson, but which are weakly bounded by electroweak
precision meassurements, i.e. we simply assume that there is some 
unspecified physics at the scale of the mass of the
new particles that makes their contribution to the Peskin-Takeuchi
parameters $S,T$ to be within experimental range \cite{peskintak}.
 The second case (Scenario-II) will include a heavy fourth family,
for which we find that the corrections induced by the heavy fermions  
on the vertices $hgg$ and $hWW/hZZ$ are correlated, and in fact can be written in 
terms of the parameter $T$. 
We also study gluon fusion for the MSSM higgs
bosons, as a possible test for the mass spectrum
of squarks. Large effects are obtained when the loop amplitude includes
non-degenerate squark masses, which could be realized in 
many particular models \cite{polonskypom}.
These results suggest that the discovery of a Higgs boson can be turned
into a new tool for high precision electroweak physics studies.

{\bf {2.- Parametrization of new physics in Higgs couplings.}}
In order to probe heavy scales through their effect on
the Higgs coupling, we shall consider extensions 
of the SM that include additional particles, but with a 
minimal Higgs sector consisting of one doublet.
We are interested in the possible representations 
that can receive their mass from the SM Higgs 
mechanism. Because of the quantum numbers of the SM Higgs, we are
restricted to consider only fermion doublets and singlets. 
Then the possibilities reduce to:
a) additional families with SM quantum numbers
b) additional mirror families
and c) a combination of the above.
Within cases a and b, new quarks must have the same color properties
as in the SM, but since for case c it is possible to cancel
anomalies among the quarks, they could lay in larger $SU(3)_c$
representations, like sextets or octets.
Since these new states have electroweak charges, they will contribute
in general to the parameters S,T,U, but not necessarily. 
Present global fits for electroweak data seem to exclude
more than one additional SM family \cite{erlerlangack}, 
however this conclusion relies
on the assumption that no other physics occurs at the energy scale
of the new fermion masses, which may not be the case in SUSY models.
 For instance, consider a model with a complete SUSY 4th family, 
in the limit $m_A >> m_Z$, under which the light Higgs boson $h^0$
behaves like the SM Higgs, and also assume a mass 
degeneracy for the components of the fermion doublets,
as well as for the sfermions, this gives $T=0$;
whereas fermions give $S=2 N_c/6\pi$, the contribution of sfermions
to $S$ is given by 
$S=\frac{N_c}{36 \pi}log(\frac{m^2_{\tilde u_{Li}} }{m^2_{\tilde d_{Li}} })$
\cite{siannah}, 
thus by choosing the appropriate masses one could decrease the total
value of S, and satisfy present experimental constraints.  
This first scenario, where we have new particles that receive their mass 
from the SM Higgs mechanism, but are weakly constrained by electroweak 
precision data, is the right place where the study of the
Higgs signal can test the presence of such heavy
particles. For the second case, when the corrections to the
different vertices $hVV$ are correlated,  one must include
all the corrections to the relevant branching ratios for the analysis of 
bounds; here we can also obtain bounds
on the heavy quanta, though they will be somewhat limited.

We shall describe the effects of heavy quanta on the higgs couplings
to gauge bosons, by writing them as: 
$g_{hVV}= g^{SM}_{hVV} (1+\epsilon_V)$,
where $V=W,Z,g,\gamma$, and similarly for the fermion couplings
$g_{hff}= g^{SM}_{hff} (1+\epsilon_f)$. The parameters $\epsilon_{f,V}$,  
encompass the effects of heavy quanta.
In general, the expression for $\epsilon_{g,\gamma}$
will be give by a ratio of complicated loop
expresions, however since we are interested in the limit
$m_h << M_{heavy}$, one can use the low-energy theorems \cite{LowETH},
to relate $\epsilon_{g,\gamma}$ to the beta-function coefficients for 
the corresponding coupling constant ($\beta_{I,X}$). 
Thus, the higgs-gluon coupling is given by
$\epsilon_g= \beta_{3,X}/\beta_{3,t}$, where 
$\beta_{3,t (X)}$ denote the contribution of top
and heavy quanta to strong beta function.
The values of $\epsilon_g$ arising from a pair of heavy color triplets, 
sextets and octet quarks  are : $\epsilon_g= 2,\,  5$ and 6, respectively,
whereas a new SM family plus its mirror partner gives $\epsilon_g= 4$. 
It is remarkable that coming stages of Tevatron
will be able to test these values at significant levels.

The correction induced by the heavy quanta
on the couplings $hWW,hZZ$ can also be derived using the
low-energy theorems. However, for the case of a fourth heavy
family, one finds an interesting relation between $\epsilon_{f,W,Z}$
and the corresponding expression for the  parameter 
$T$, namely, 
\begin{eqnarray}
\epsilon_f&=& \frac{\alpha T}{2}+ \frac{N_c G_F }{12\sqrt{2} \pi^2 } 
                    (m^2_1+m^2_2) \ \nonumber \\
\epsilon_Z&=& -\frac{N_c G_F }{6 \sqrt{2} \pi^2 }  (m^2_1+m^2_2) \nonumber \\
\epsilon_W&=& \epsilon_Z + \frac{\alpha T }{2}
\end{eqnarray}
where $T$ includes the contribution of the fourth family fermions with masses
$m_{1,2}$.

We could also consider effects arising from heavy particles  
that receive their mass from the breaking of new symmetries 
characterized by a large scale $V_{new} >> v=246$ GeV, induced by another 
set of heavy Higgs bosons ($\Phi_{new}$). Supposse this
is communicated to the SM Higgs by including in the scalar potential
a mixing term  of the form:  $\lambda |\Phi_{new}|^2 |\phi_{sm}|^2$.
Then one can also include the contribution of these ultra-heavy
particles to the parameters $\epsilon_V$.
However, we found that in this case the mixing angle 
that transforms the Higgs weak to mass-eigenstate  basis,
is suppressed by the large scale ($V_{new}$), and it induces
decoupling effects at low-energies, namely 
$\epsilon_g \simeq v/V_{new}$.

{\bf {3.- The mechanism of gluon fusion and bounds on $\epsilon_g$ .}} 
The contribution coming from the new particles
will modify the cross section for gluon fusion,
as follows:
$\sigma (pp\to h+X) = \sigma_{SM} (1+ \epsilon_g)^2$
where the SM cross-section
$\sigma_{SM}$ can be written in terms of the Higgs
decay width $\Gamma(h_{SM} \to gg)$; details of the
equations can be found in the literature \cite{hspira},
Since the new particles modify the decay width into $h\to gg$, 
it will enhance the production by gluon fusion reaction, 
but this enhancement could work against the final signal rate, 
which usually involves some decay of the Higgs boson into a mode
with clear signature, like $h\to WW^*$ or $\gamma \gamma$ 
for the intermediate
Higgs mass region. Both effects can be taken properly into account
in the analysis by writing the product of the cross-section
times the branching ratio of the signal as
$[\sigma \times B.R.(h\to VV)]_{new}= 
R_V \times [\sigma \times B.R.(h\to VV)]_{SM}$,
where $R_V$ is given by ,
\begin{equation}
R_V  =   \frac{(1+ \epsilon_g)^2(1+\epsilon_V)^2 }
       {[ 1+\sum_Y (\epsilon^2_Y + 2 \epsilon_Y)*B.R.(h_{SM} \to YY)]}
\end{equation}  
where the sum in the denominator runs over $Y=t,b,g,A,W,Z$.

We can obtain bounds on the parameters $\epsilon_{g,W}$ using gluon fusion
and the decay $h\to WW^*$ at Tevatron, which was studied in  detail in 
ref. \cite{Hanetal}, this work concluded that it is possible to
detect a SM Higgs boson with an integrated luminosity of 30 fb$^{-1}$,
provided that an optimized selection of cuts is implemented; here we shall
only use their first stage of cuts, namely: 
transverse lepton momentum $p_t(e,\mu)>10$ GeV, 
psedorapidity $\eta_{e,\mu} < 1.5$,
lepton invariant mass $m_{ll}>10$ GeV,
jet resolution $\Delta R(l-j) >0.4$,
missing transverse energy $E_T > 10$ GeV.
For the case when the corrections to $\epsilon_{W}$ can be neglected
(Scenario-I), the above cuts already allow us to get interesting bounds 
on the parameter $\epsilon_g$ at  95 \% c.l., 
as it is shown in figure 1 ( for 2 and 10 $fb^{-1}$). 
These bounds will in turn constrain the
presence of heavy colored particles coming from cases
a, b and c, provided that the Higgs mass lays in the
intermediate mass range. The interception of the exclusion lines
with the straight dashed line (which corresponds to 
$\epsilon_g=2$), shows the values of Higgs
masses where it will be possible to limit the presence of a
pair of heavy color triplets. Furthermore, since 
sextets and octet  quarks give a larger value of $\epsilon_g$,
their presence could be excluded too.

However, these bounds need some clarfication for the case when 
the new physics also modifies the parameters $\epsilon_{W,Z}$
, since this will
affect the decay rates into $h\to WW^*$. For the fourth
family case (Scenario-II), the corrections to the vertex $hVV$ are negative
(as compared with the positive tree-level value)
and grow with the mass of the heavy quanta; 
for instance for fermion masses of order 500 (700) GeV 
the deviations from the SM tree-level
couplings are of order - 27 (-52) \% \cite{veltmanetal}, 
which seems to imply that the signal will no longer be detectable
i.e. one can only probe heavy but not ultraheavy quanta.
 With 10 $fb^{-1}$ of integrated luminosity,
it will be possible to exclude only a limited range of
fermion masses, up to about 900 GeV at best, 
as it is shown in table 1.
However since for these heavy masses one enters into
the non-perturbative domain one can not draw
a definite  conclusion, unless it is found an scheme to treat
the non-perturbative Higgs effects.
On the other hand,  the decay into photon pair
can also be used to probe higgs couplings at LHC; but in this case
the W-loop dominates the amplitude, and the contribution from
addditional fermions has
opposite sign, which will make it difficult to separate the effects,
or even worst, may conspire to reduce the signal.

{\bf {3.- Probing a Non-universal sfermion spectrum.}}
The MSSM includes two Higgs doublets,  and the
Higgs spectrum consists of two neutral CP-even scalars $h^0$ and $H^0$,
one CP-odd pseudoscalar $A^0$ and a charged pair $H^\pm$ \cite{himssm}.
The Higgs sector of the model is completely determined at tree level by 
fixing two parameters, conventionally chosen to be $\tan\beta$ 
and the pseudoscalar mass $m_A$. At loop levels, the radiative corrections,
mainly from top and  stop, 
modify the tree-level mass bound ($m_h < m_Z \cos 2\beta$),
allowing to have  $m^{max}_h \simeq 130$ GeV \cite{carenaetal}. 
The corrections to Higgs couplings can also lead to
modification of its production mechanisms 
at hadron colliders, as previously studied 
\cite{adjouadi,tlcgroup}. 
Although conventional wisdom states that the contribution of
top/bottom quarks (squarks)  dominate the amplitude for the gluon fusion,
and squarks from first and second generations can be neglected, 
it should be mentioned that this result only holds within the minimal 
SUSY breaking scenarios, where it is typically
assumed that sfermions are mass degenerate,
as required  to respect FCNC constraints \cite{susyfcnc}.
 However, mass non-degeneracy for sfermions within the same family 
but different isospin can be acceptable, 
since it will only be midly constrained by the parameter $T$. 
In fact, some  non-degeneracy can arises even in 
the minimal SUSY-GUT, by imposing universal boundary  conditions  
at the Planck scale, rather than the GUT scale. 
Other cases that require non-degeneracy are
models with additional D-terms \cite{polonskypom},
and in Finite Grand Unification \cite{finitegut}.

Thus, it is interesting to find some process that could 
test the sfermion mass spectrum. In this letter we consider
the decay $h \to gg$ (which determines the gluon fusion
mechanism) to probe the structure of soft-breaking masses predicted by 
models of SUSY breaking.
The expression for the decay width of $h\to gg$ that includes
a non-universal mass spectrum for squarks, 
can be written  as,
\begin{equation}
\Gamma(h\to gg)= \frac{G_F \alpha^2_s m^3_h }{ 4\sqrt{2} \pi }|G_{h}|^2
\end{equation}
where
\begin{equation}
G_{h^0}= G_{t,{\tilde t}} + G_{b,{\tilde b}} + G_{LR}
          + G_{UD}
\end{equation}
For $h=h^0$, $G_h$ is given by:
\begin{equation}
G_{t,{\tilde t}} = - \frac{m^2_t}{m^2_h} \frac{\cos\alpha}{\sin\beta} 
 [ f_1(\lambda_t)+ f_3(\lambda_{\tilde t_L})+ f_3(\lambda_{\tilde t_R})] 
\end{equation}
\begin{equation}
G_{b,{\tilde b} }= \frac{m^2_b}{m^2_h} \frac{\sin\alpha}{\cos\beta} 
 [ f_1(\lambda_b)+ f_3(\lambda_{\tilde b_L})+ f_3(\lambda_{\tilde b_R})] 
\end{equation}
\begin{eqnarray}
G_{LR}&=&  \frac{m^2_W s^2_w}{m^2_h c^2_W} \sin(\beta+\alpha) 
                   [Q_u (f_3(\lambda_{\tilde u_R})- 
                  f_3(\lambda_{\tilde u_L}) \nonumber \\
            & &+   Q_d (f_3(\lambda_{\tilde d_R})- f_3(\lambda_{\tilde d_L}) ] 
\end{eqnarray}
\begin{equation}
 G_{UD} =  \frac{m^2_W }{2 m^2_h c^2_W} \sin(\beta+\alpha) 
\Sigma_{u,d} [f_3(\lambda_{\tilde u_L}) - f_3(\lambda_{\tilde d_L}) ] 
\end{equation}
The expression for $h=H^0$ are obtained by making the
replacements $\cos\alpha \to -\sin\alpha$, $\sin\alpha \to -\cos\alpha$, 
$\sin(\alpha+\beta) \to -\cos(\alpha+\beta)$;
the explicit form of $f_1(z),f_3(z)$ can be found in ref.
\cite{hspira}.
The contribution proportional to the 
fermion masses are kept only for the stop and sbottom 
($G_{t,{\tilde t} }, G_{b,{\tilde b} } $),
whereas the remmaining SUSY-breaking effects are kept for all sfermions
($G_{LR}, G_{UD}$). To reduce the number
of parameters, we choose our $\mu, A_t$ values in such a way that one can
neglect L-R mixing, which was found previously to
give small effects\cite{adjouadi}.
We have used the previous equation to evaluate the contribution of 
squarks to the parameter $\epsilon_g$, for
 $m_A=200$ GeV, $\tan\beta=5,20$, and
squark masses covering the range from 200 to 700 GeV.
For simplicity,
we apply the 1-loop leading log formula to the Higgs masses and parameters.
Results are shown in table 2, and for comparision
we have also included the values obtained with universal squark masses.
These results show that the effect of non-universal masses  
can modify the $B.R.(h\to gg)$, by values of order
$\pm 20 \% $.

{\bf {4.- Conclusions.}}
We have studied the Higgs interaction with fermions and gauge bosons, 
and found that deviations from the SM prediction, 
induced by heavy quanta, can be described by a set of parameters 
$\epsilon_{f,V}$, which can be tested at future colliders.
The production cross-section by gluon fusion 
can be used to probe $\epsilon_g$ at Tevatron, whereas the promising
Higgs decays into $ WW^*$ can probe $\epsilon_{W}$. 
We find that the resulting bounds may imply the exclusion of heavy 
particles that receive their mass directly from the SM Higgs, 
including  a 4th standard or mirror family or chiral colored sextets and octets. 
For the 4th family case we disply a relations between
$\epsilon_{f,V}$ and the Peskin-Takeuchi parameter $T$.
Within the MSSM, we also find that gluon fusion is a sensitive
probe for a non-universal spectrum of 
squarks masses, 
which allows for the possibility to test 
the sfermion masses predicted in
 the models of SUSY breaking.
Whether the discovery of a Higgs signal  
will lead Particle Physics into a 
dark middle-age or into a  renaissance remains to be seen, 
but the answer will depend 
crucially on our hability to meassure 
the Higgs properties, which could reveal 
us the path to the promised land of new physics 
or may be just the boring 
solitude of the SM up to very heavy scales. 

\smallskip
{\small 
We thank M. Chanowitz and C. Kolda for useful discussions.
The support from ``Fundacion Mexico-USA para la ciencia''
and the hospitality of LBL are also acknwoledged.
JLDC is also supported by CONACYT-SNI (Mexico). }




{\bf FIGURE CAPTION}

\bigskip

Fig 1. 95 \% exclusion contours for the parameter $\epsilon_g$
that can be obtained from Higgs search at Tevatron RUN-II,
with  2 (dashes) and 10 (solid) $fb^{-1}$ of integrated luminosity.
The straight (dashed) lines corresponds to $\epsilon_g=2$.
\bigskip
{\bf TABLE CAPTION}

\bigskip

Table 1. Upper limits to 4th generation fermion masses
that can be obtained from Higgs search at Tevatron RUN-II,
with  10 $fb^{-1}$.

\bigskip

Table 2. Values predicted for the parameter $\epsilon_g$
for the MSSM with $\tan\beta = 5,20$. $M_{\tilde Q}$
represents the range of values taken for
$M_{\tilde U_R}$, $M_{\tilde D_L}$,  $M_{\tilde D_R}$. 

\bigskip

\bigskip

{\bf \, \, \, Table. 1} 

\begin{center}
\begin{tabular}{||l| l||}
\hline
$m_h$ [GeV]  & $m^{max}_{4th}$ [GeV]\\
\hline 
\,\, 120. & \,\, 350. \\
\hline
\,\, 130. & \,\, 520. \\
\hline
\,\, 140. & \,\, 650. \\
\hline
\,\, 150. & \,\, 730. \\
\hline
\,\,  160. & \,\, 870. \\
\hline
\,\, 170. & \,\, 710. \\
\hline
\,\, 180. &\,\, 670. \\
\hline
\,\, 190. & \,\, 600. \\
\hline
\,\,  200. & \,\, 510. \\
\hline
\end{tabular}
\end{center}

\bigskip

\bigskip

{\bf \, \, \, Table. 2} 

\begin{center}
\begin{tabular}{|| l | l | l | l ||}
\hline
$m_{\tilde U}$ [GeV]  & $m_{\tilde Q}$ [GeV]  & $\epsilon_g$ (non-univ) & 
$\epsilon_g$ (univ) \\
\hline 
 200. & $200\to 700$ & \, $0.1 \to 0.4$  &  0.4  \\
\hline
 500. &  $200\to 700$ & $-0.05 \to 0.14$  &  0.05  \\
\hline
 700. &  $200\to 700$ & $-0.05 \to 0.18$  &  0.1  \\
\hline
\end{tabular}
\end{center}

\bigskip


\begin{references}


\vspace*{-1.4cm}

\bibitem{Higgsrev} For a recent review of higgs mass bounds
see: M. Kado, talk at Recontres de Moriond (March, 2000),
hep-ex/0005022.

\bibitem{Meandthem} J.L Diaz-Cruz et al., Mod. Phys. Lett. A15 (2000)
1377 [hep-ph/9905335];
C. Kolda and L. Hall, Phys. Lett. B459 (1999) 213
[hep-ph/9904236].

\bibitem{carenaetal} For an updated calculation and references
to earlier work see: M. Carena et al., hep-ph/0001002.

\bibitem{hixhunter} J. Gunion, H. Haber, G. Kane and S. Dawson,
'The Higgs Hunters Guide', Adison Wesley, Reading 1990.

\bibitem{chanofourth} M. Chanowitz, Phys. Rev. Lett. 69 (1992) 2037;
I.F. Ginzburg, I.P. Ivanov and A. Shiller, Phys. Rev. D60 (1999) 095001.

\bibitem{veltmanetal} M. Veltman, Nucl. Phys. B123 (1977) 89;
M. Chanowitz, M. Furman and I. Hinchliffe, Nucl. Phys. B153 (1979) 402.


\bibitem{polonskypom} N. Polonski and A. Pomarol, Phys. Rev. D51 (1995) 6532;
C.Kolda and S. Martin, Phys. Rev. D53 (1996) 3871.

\bibitem{erlerlangack} J. Erler and P. Langacker, in Review of
Particle Properties, Eur. Phys. J C3 (1998) 1.

\bibitem{siannah} A. Dobado, M.J. Herrero and S. Penaranda,
Eur. Phys. J.  C7 (1999) 313.

\bibitem{LowETH} A. Vainshtein et al. Sov. J. Nucl. Phys. 30 (1979) 711;
B. Kniehl and  M. Spira Z. Phys. C69 (1995) 77.


\bibitem{hspira} For a review see: M. Spira, Fortsch. Phys. 46 (1998) 203. 

\bibitem{peskintak} M. Peskin and T. Takeuchi, Phys. Rev. Lett. 65 (1990) 964.


\bibitem{Hanetal} T. Han, A. Turcot and R.J. Zhang, hep-ph/9812275.


\bibitem{himssm}J. Gunion and H. Haber, Nucl. Phys. B272 (1986) 1.

\bibitem{adjouadi} A. Djouadi, Phys. Lett. B435 (1998) 101.

\bibitem{tlcgroup} J.L. Diaz-Cruz et al., Phys. Rev. Lett. 80 (1998) 4641;
C. Balaaz et al., Phys. Rev. D59 (1999) 055016; M. Carena, S. Mrenna and 
C. Wagner, Phys. Rev. D60 (1999) 075010.

\bibitem{susyfcnc} A. Masiero et al., Nucl. Phys. B477 (1996) 321.

\bibitem{finitegut} For a review see: T. Kobayashi, J. Kubo, M. Mondragon 
and G. Zoupanos, in Proccedings of Mexican School of Particles and
Fields (Morelia, Mexico, 1997).

\end{references}
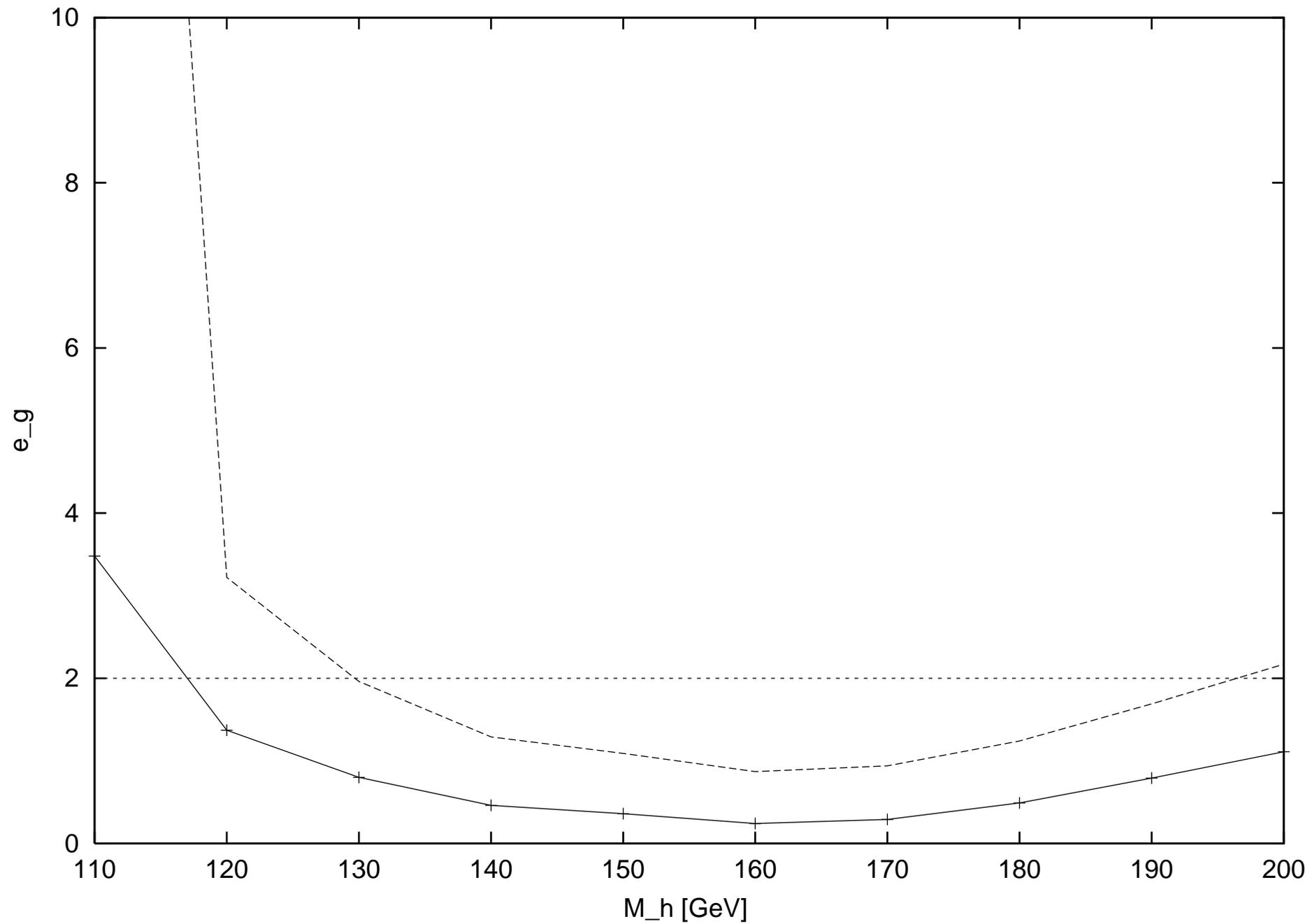
\end{document}